\documentstyle[preprint,aps,epsfig]{revtex}
\newcommand{\eq}{\begin{eqnarray}}
\newcommand{\en}{\end{eqnarray}}

\tightenlines

\begin{document}

\title{Strange nucleon form factors in \\
the perturbative chiral quark model}
\author{V. \ E. \ Lyubovitskij,  P. \  Wang,
Th. \ Gutsche, and Amand \, Faessler}
\address{Institut f\"ur Theoretische Physik, Universit\"at
T\"ubingen, Auf der Morgenstelle 14,  \\
D-72076 T\"ubingen, Germany}

\maketitle

\vskip.5cm
   
\begin{abstract}
We apply the perturbative chiral quark model at one loop
to calculate the strange form factors of the nucleon. 
A detailed numerical analysis of the strange magnetic moments 
and radii of the nucleon, and also the momentum dependence 
of the form factors is presented.
\end{abstract}

\vskip.5cm


\vskip.5cm


\section{Introduction}

Strange quark contributions to the properties of the nucleon have 
attracted a lot of interest since the originally puzzling EMC results of 
the proton spin \cite{EMC}. Recently, the SAMPLE \cite{Sample1,Sample2} 
and HAPPEX \cite{Happex1,Happex2} collaborations reported first results 
for the strange nucleon form factors from the measurement of the 
parity-violating asymmetry in elastic electron-proton scattering.
The SAMPLE collaboration at MIT-Bates \cite{Sample1,Sample2}
concentrates on the strange magnetic form factor $G_M^{\rm s}$ at a
small momentum transfer with $Q_{\rm S}^2 = 0.1$ GeV$^2$. 
The updated value of $G_M^{\rm s}$ measured by the SAMPLE 
Collaboration \cite{Sample2} is
\eq
G^{\rm s}_{\rm SAMPLE}(Q_{\rm S}^2) \doteq G_M^{\rm s}(Q_{\rm S}^2)
= 0.14 \pm 0.29 \pm 0.31.
\en
Here and following results for the magnetic form factor are given in 
units of nuclear magnetons. The HAPPEX collaboration at
TJNAF \cite{Happex1,Happex2} extracted from the data the combination 
of charge $G^{\rm s}_E$ and magnetic $G^{\rm s}_M$ form factor
\begin{equation}
G^{\rm s}_{\rm HAPPEX}(Q_{\rm H}^2) \doteq G^{\rm s}_E(Q_{\rm H}^2) +
0.39 G^{\rm s}_M(Q_{\rm H}^2) =  0.025 \pm 0.020 \pm 0.014
\end{equation}
at $Q_{\rm H}^2=0.477$ GeV$^2$ \cite{Happex2}. The upcoming experiment
by the A4 Collaboration at MAMI \cite{Koebis} intends to measure
\begin{equation}
G^{\rm s}_{\rm MAMI}(Q_{\rm M}^2) \doteq G^{\rm s}_E(Q_{\rm M}^2) +
0.22 G^{\rm s}_M(Q_{\rm M}^2),
\end{equation}
where $Q_{\rm M}^2=0.23$ GeV$^2$.

Different theoretical approaches were applied to the analysis of  
strange nucleon form factors, including QCD equalities and Lattice 
QCD \cite{Leinweber_QCD,Leinweber,Dong,Mathur}, heavy baryon chiral 
perturbation theory (HBChPT) \cite{Hemmert,Hemmert2}, dispersive  
approaches \cite{Jaffe}-\cite{Hammer}, kaon loop  
calculations \cite{Musolf}, a hybrid model based on vector meson 
dominance (VMD) in addition to a kaon cloud contribution \cite{Cohen}, 
VMD model \cite{Dubnicka}, Skyrme \cite{Park}, NJL \cite{Weigel} and 
chiral \cite{Silva} soliton models, chiral bag \cite{Hong} and chiral 
quark \cite{Glozman,Hannelius} models, etc. Theoretical results vary 
quite widely. For example, the predictions for the strange magnetic  
moment $\mu_N^s \doteq G_M^S(0)$ are distributed from negative values  
$- 0.75\pm 0.30$ \cite{Leinweber_QCD} to positive ones $0.37$  
\cite{Hong}. Theoretical results for the strange charge radius  
vary from $- 0.16 \pm 0.06$~fm$^2$ \cite{Dong} to $0.05 \pm 0.09$~fm$^2$ 
\cite{Hemmert2}.  Similarly, predictions for the momentum dependence of 
the strange form factors cover a wide range of values, depending on the 
individual model. Therefore, more precise experiments can serve as 
a crucial check for existing low-energy approaches, which describe 
the sea of strange quarks inside the nucleon. 

Here we concentrate on the calculation of the strange nucleon form 
factors in the framework of the perturbative chiral quark model (PCQM) 
which was suggested and developed in \cite{Gutsche}-\cite{PCQM3} for 
the study of low-energy properties of baryons. The PCQM is based on the 
nonlinear $\sigma$-model quark Lagrangian and includes a 
phenomenological confinement potential. Baryons are considered as bound 
states of valence quarks surrounded by a cloud of pseudoscalar mesons. 
We treat the meson cloud perturbatively and, therefore, our approach is 
similar to quark models studied in \cite{Hannelius}, 
\cite{Theberge}-\cite{Chin}. The model was successfully applied to the 
electromagnetic properties of the nucleon \cite{PCQM1}, $\sigma$-term 
physics \cite{PCQM2} and the $\pi N$ scattering including radiative 
corrections \cite{PCQM3}.

In the present article we proceed as follows. In Sec. II we briefly
describe the basic notions of our approach. In Sec. III we apply the
model to investigate the strange form factors of nucleon. Sec. IV
contains a summary of our major conclusions.

\section{Perturbative chiral quark model}

The starting point of the perturbative chiral quark model
(PCQM) \cite{PCQM1}-\cite{PCQM3} is an effective chiral Lagrangian
describing the valence quarks of baryons as relativistic fermions
moving in a self-consistent field (static potential)
$V_{\rm eff}(r)=S(r)+\gamma^0 V(r)$ with
$r=|\vec{x}|$ \cite{Gutsche,Oset}, which are supplemented by a cloud of
Goldstone bosons $(\pi, K, \eta)$. (For details see
Refs. \cite{PCQM1}.) The effective Lagrangian 
${\cal L}_{\rm eff} = {\cal L}_{inv} + {\cal L}_{\chi SB}$ of the PCQM 
includes a chiral invariant part ${\cal L}_{\rm inv}$ and a symmetry 
breaking term ${\cal L}_{\chi SB}$ (containing the mass terms for quarks 
and mesons):
\begin{eqnarray}
{\cal L}_{\rm inv}(x)&=&\bar\psi(x) [i \not\!\partial - \gamma ^{0}V(r)]
\psi(x) + \frac{1}{2}[D_{\mu }\Phi_i(x)]^{2} - S(r) \bar\psi(x)
\exp \left[i\gamma^{5} \frac{\hat{\Phi}(x)}{F}\right]\psi (x), \\
{\cal L}_{\chi SB}(x)&=&-\bar\psi(x){\cal M}\psi(x)
-\frac{B}{2} {\rm Tr} \biggl[ \hat\Phi^2(x) \, {\cal M} \biggr],
\end{eqnarray}
where $\hat{\Phi}(x)$ is the octet matrix of pseudoscalar mesons,
$D_\mu$ is the covariant derivative \cite{PCQM2};
$F = 88$ MeV \cite{PCQM1,Gasser} is the pion decay constant in the chiral
limit; ${\cal M} = {\rm diag}\{\hat m, \hat m, m_s\}$ is the mass matrix
of current quarks with $\hat m = 7$ MeV and $m_s = 25 \hat m$;
$B = 1.4$ GeV is the quark condensate constant\footnote{Here we
restrict to the isospin symmetry limit with $m_u = m_d = \hat m$.}.
We rely on the standard picture of chiral symmetry
breaking~\cite{Gasser_Leutwyler} and for the masses of pseudoscalar
mesons we use the leading term in their chiral expansion (i.e. linear
in the current quark mass):
\begin{eqnarray}\label{M_Masses}
M_{\pi}^2=2 \hat m B, \hspace*{.5cm} M_{K}^2=(\hat m + m_s) B,
\hspace*{.5cm} M_{\eta}^2= \frac{2}{3} (\hat m + 2m_s) B.
\end{eqnarray}
We expand the quark field $\psi$ in the basis of potential eigenstates as
\begin{eqnarray}\label{total_psi}
\psi(x)&=&\sum\limits_\alpha b_\alpha u_\alpha(\vec{x})
\exp(-i{\cal E}_\alpha t) + \sum\limits_\beta d_\beta^\dagger
v_\beta(\vec{x}) \exp(i{\cal E}_\beta t),
\end{eqnarray}
where the sets of quark $\{ u_\alpha \}$ and antiquark $\{ v_\beta \}$
wave functions in orbits $\alpha$ and $\beta$ are solutions of the
Dirac equation with the static potential. The expansion coefficients
$b_\alpha$ and $d_\beta^\dagger$ are the corresponding single quark
annihilation and antiquark creation operators.

Treating Goldstone fields as small fluctuations around the three-quark
(3q) core we formulate perturbation theory in the expansion parameter
$1/F$ ($F \sim \sqrt{N_c})$ and we also treat finite current quark masses
perturbatively \cite{PCQM1}. In this paper we restrict to the linear 
form of the meson-quark interaction $\exp [i\gamma^{5} \hat\Phi(x)/F] 
\approx 1 + i\gamma^{5} \hat\Phi(x)/F$ and all calculations are 
performed at one loop or at order of accuracy $o(1/F^2, \hat{m}, m_s)$. 
In the calculation of matrix elements we project quark diagrams on the 
respective baryon states. 
The baryon states are conventionally set up by the product of the 
${\rm SU(6)}$ spin-flavor and ${\rm SU(3)_c}$ color wave functions, 
where the nonrelativistic single quark spin wave function is replaced by 
the relativistic solution $u_\alpha(\vec{x})$ of the Dirac equation
\begin{equation}
\left[ -i\gamma^0\vec{\gamma}\cdot\vec{\nabla} + \gamma^0 S(r) + V(r)
- {\cal E}_\alpha \right] u_\alpha(\vec{x})=0,
\end{equation}
where ${\cal E}_\alpha$ is the single-quark energy.

For the description of baryon properties we use the effective potential
$V_{\rm eff}(r)$ with a quadratic radial dependence \cite{PCQM1,PCQM2}:
\eq
S(r) = M_1 + c_1 r^2, \hspace*{1cm} V(r) = M_2+ c_2 r^2
\en
with the particular choice
\eq
M_1 = \frac{1 \, - \, 3\rho^2}{2 \, \rho R} , \hspace*{1cm}
M_2 = {\cal E}_0 - \frac{1 \, + \, 3\rho^2}{2 \, \rho R} , \hspace*{1cm}
c_1 \equiv c_2 =  \frac{\rho}{2R^3} .
\en
Here, ${\cal E}_0$ is the single-quark ground-state energy;
$R$ are $\rho$ are parameters related to the ground-state quark wave
function $u_0$:
\begin{eqnarray}\label{Gaussian_Ansatz}
u_0(\vec{x}) \, = \, N \, \exp\biggl[-\frac{\vec{x}^{\, 2}}{2R^2}\biggr]
\, \left(
\begin{array}{c}
1\\
i \rho \, \vec{\sigma}\vec{x}/R\\
\end{array}
\right)
\, \chi_s \, \chi_f \, \chi_c,
\end{eqnarray}
where $N=[\pi^{3/2} R^3 (1+3\rho^2/2)]^{-1/2}$ is a normalization
constant; $\chi_s$, $\chi_f$, $\chi_c$ are the spin, flavor and color
quark wave function, respectively. Note, that the constant part of the 
scalar potential $M_1$ can be interpreted as the constituent mass of 
the quark, which is simply the displacement of the current quark mass 
due to the potential $S(r)$. The parameter $\rho$ is related to the 
axial charge $g_A$ of the nucleon calculated in zeroth-order 
(or 3q-core) approximation:
\begin{eqnarray}\label{ga_rho_match}
g_A=\frac{5}{3} \biggl(1 - \frac{2\rho^2} {1+\frac{3}{2}\rho^2}\biggr) .
\end{eqnarray}
Therefore, $\rho$ can be replaced by $g_A$ using the matching condition
(\ref{ga_rho_match}). The parameter $R$ is related to the charge radius
of the proton in the zeroth-order approximation as
\begin{eqnarray}\label{rad_LO} 
<r^2_E>^P_{LO} = \int d^3 x \, u^\dagger_0 (\vec{x}) \,
\vec{x}^{\, 2} \, u_0(\vec{x}) \, = \, \frac{3R^2}{2} \,
\frac{1 \, + \, \frac{5}{2} \, \rho^2}{1 \, + \, \frac{3}{2} \, \rho^2}.
\end{eqnarray}
In our calculations we use the value $g_A$=1.25 \cite{PCQM1}. Therefore,
we have only one free parameter, that is $R$. In the numerical
studies \cite{PCQM1} R is varied in the region from 0.55 fm to 0.65 fm,
which corresponds to a change of $<r^2_E>^P_{LO}$  
from 0.5 to 0.7 fm$^2$.

\section{Strange form factors of nucleon}

Now we consider the calculation of the strange vector and axial vector 
nucleon form factors. We first derive from the model Lagrangian the 
strangeness vector current $V^{\rm s}_\mu$, which is the combination of 
the baryonic $J^{\rm B}_\mu$ and the hypercharge $J^{\rm Y}_\mu$ currents:
\eq
V^{\rm s}_\mu \, \doteq \,  J^{\rm B}_\mu - J^{\rm Y}_\mu \,\, .
\en
Using Noether's theorem we have:
\eq
J^{\rm B}_\mu \, = \, \frac{1}{3} \, \bar q \, \gamma_\mu \, q ,
\hspace*{1cm}
J^{\rm Y}_\mu \, = \, \bar q \, \gamma_\mu  \,
\frac{\lambda_8}{\sqrt{3}} \, q \, + \,
\frac{2}{\sqrt{3}} \, f_{8ij} \, \Phi_i \, \partial_\mu \, \Phi_j,
\en
where $f_{ijk}$ are the totally antisymmetric structure constants of
${\rm SU(3)}$. We therefore obtain: 
\eq\label{Vs_current}
V^{\rm s}_\mu \, = \, \bar s \gamma_\mu s  +  (K^+ i \partial_\mu
K^-  + K^0 i \partial_\mu  \bar K^0  + {\rm h.c.}) \,\,\, .
\en
Note, that both valence and sea ($K$-meson cloud) quarks contribute to
the strangeness vector current. This form of the strangeness current is
common to chiral quark models \cite{Hannelius}.
The strangeness axial current $A^{\rm s}_\mu$ is given by
\eq\label{As_current}
A^{\rm s}_\mu \, = \, \bar s \gamma_\mu \gamma_5 s  \,\, , 
\en
where, because of the pseudoscalar nature of the Goldstone bosons, the 
mesonic piece is absent. To perform a consistent calculation of strange 
nucleon form factors we have to guarantee local gauge invariance associated 
with the electromagnetic $U_{\rm em}(1)$ group. As in the case of 
electromagnetic form factors (see Ref. \cite{PCQM1}), we restrict our 
kinematics to a specific frame, that is the Breit frame (BF), where gauge 
invariance is fulfilled due to the decoupling of the time and vector 
components of the electromagnetic current operator \cite{Miller-Thomas}. 
In the BF the initial momentum of the nucleon is $p = (E, -\vec{q}/2)$, 
the final momentum is $p^\prime = (E, \vec{q}/2)$ and the 4-momentum-transfer 
is $q = (0, \vec{q}\,)$ with $p^\prime = p + q$. With the space-like
momentum transfer squared given as $Q^2 = - q^2 = \vec{q}^{\, 2}$,  
the strange form factors of nucleon -  charge $G_E^{\rm s}$ and
magnetic $G_M^{\rm s}$ (Sachs) form factors - are defined in the BF by
\begin{eqnarray}
& &<N\biggl(\frac{\vec{q}}{2}\biggr)|V^{\rm s}_0(0)|
N\biggl(-\frac{\vec{q}}{2}\biggr)>
= \chi^\dagger_N \chi_N G_E^{\rm s}(Q^2) , \\
& &<N\biggl(\frac{\vec{q}}{2}\biggr)|\vec{V}^{\rm s}(0)|
N\biggl(-\frac{\vec{q}}{2}\biggr)>
= \chi^\dagger_N \frac{i \vec{\sigma}_N \times \vec{q}}{2m_N}
\chi_N G_M^{\rm s}(Q^2) ,  \\
& &<N\biggl(\frac{\vec{q}}{2}\biggr)| A^{\rm s}_3(0)|
N\biggl(-\frac{\vec{q}}{2}\biggr)>
= \chi^\dagger_N \sigma_N^3 \chi_N G_A^{\rm s}(Q^2) .
\end{eqnarray}
Here, $V^{\rm s}_0$ and $\vec{V}^{\rm s}$ are the time and space
component of the strangeness vector current~(\ref{Vs_current}); 
$A^{\rm s}_3$ is the third spatial component of the strangeness axial 
current~(\ref{As_current}); $\chi_N$ is the nucleon spin wave function. 
At zero recoil the Sachs form factors satisfy the following 
normalization conditions
\begin{eqnarray}
G_E^{\rm s}(0)=0, \hspace*{1cm} G_M^{\rm s}(0)=\mu^{\rm s},
\hspace*{1cm} G_A^{\rm s}(0)=g_A^{\rm s} \, ,
\end{eqnarray}
where $\mu^{\rm s}$ and $g^{\rm s}_A$ are the strange nucleon
magnetic moment and axial charge, respectively. 
Dirac $F_1^{\rm s}(Q^2)$ and Pauli $F_2^{\rm s}(Q^2)$ form factors
are related to the Sachs form factors by
\eq
G_E^{\rm s}(Q^2) &=& F_1^{\rm s}(Q^2) - \frac{Q^2}{4m_N^2}
F_2^{\rm s}(Q^2) \, , \\
G_M^{\rm s}(Q^2) &=& F_1^{\rm s}(Q^2) + F_2^{\rm s}(Q^2) \, .
\en
The strange nucleon radii (charge, magnetic and axial) are given by
\begin{eqnarray}
<r^2>^{\rm s}_{\rm I} = - 6 \frac{dG^{\rm s}_{\rm I}(Q^2)}{dQ^2}
\Bigg|_{\displaystyle{Q^2 = 0}} \,\, , \hspace*{1.5cm} {\rm I} = E, M, A
 \,\, .
\end{eqnarray}

In the PCQM the strange vector (charge and magnetic) and axial form
factors of the nucleon are defined by
\begin{eqnarray}
\chi^\dagger_N \chi_N G_E^{\rm s}(Q^2) &=& \,
^N\!\!<\phi_0| \, - \frac{1}{2} \,
\int \, \delta(t) \, d^4x \, d^4x_1 \, d^4x_2 \, e^{-iqx} \,
\nonumber\\
&\times& T[{\cal L}_{int}(x_1) \, {\cal L}_{int}(x_2) \,
V^{\rm s}_0(x)] \, |\phi_0>_{c}^{N} , \label{GES}\\
\chi^\dagger_N\frac{i \vec{\sigma}_N\times\vec{q}}{2m_N}\chi_N
G_M^{\rm s}(Q^2)
&=&   \, ^N\!\!<\phi_0| \, - \frac{1}{2} \,  \int \, \delta(t) \,
d^4x \, d^4x_1 \, d^4x_2 \, e^{-iqx} \, \nonumber\\
&\times& T[{\cal L}_{int}(x_1) \, {\cal L}_{int}(x_2) \,
\vec{V}^{\rm s}(x)] \, |\phi_0>_{c}^{N} , \label{GMS} \\
\chi^\dagger_N \sigma^3_N \chi_N G_A^{\rm s}(Q^2)
&=&   \, ^N\!\!<\phi_0| \, - \frac{1}{2} \,
\int \, \delta(t) \, d^4x \, d^4x_1 \,
d^4x_2 \, e^{-iqx} \, \nonumber \\
&\times& T[{\cal L}_{int}(x_1) \, {\cal L}_{int}(x_2) \,
A^{\rm s}_3(x)] \, |\phi_0>_{c}^{N} \, \label{GMA} ,
\end{eqnarray}
where ${\cal L}_{int}$ is the linearized strong interaction Lagrangian
of kaons and quarks:
\begin{eqnarray}\label{L_I}
{\cal L}_{int}(x) = - \frac{S(r)}{F} \biggl[
K^+(x) \bar u(x) i\gamma^5 s(x) +
K^0(x) \bar d(x) i\gamma^5 s(x) \biggr] \, + \, h. c. 
\end{eqnarray}
Superscript $"N"$ in Eqs. (\ref{GES}) and (\ref{GMS}) indicates that the
matrix elements are projected on the respective nucleon states and
subscript $"c"$ refers to contributions from connected graphs only.
At one loop the relevant diagrams are indicated by Figs.1a (meson-cloud) 
and 1b (vertex correction). For the quark field we use a Feynman
propagator for a fermion in a binding potential:
\begin{eqnarray}\label{quark_propagator}
i G_\psi(x,y) &=& \theta(x_0-y_0) \sum\limits_{\alpha} u_\alpha(\vec{x})
\bar u_\alpha(\vec{y}) e^{-i{\cal E}_\alpha (x_0-y_0)}
- \theta(y_0-x_0) \sum\limits_{\beta} v_\beta(\vec{x})
\bar v_\beta(\vec{y}) e^{i{\cal E}_\beta (x_0-y_0)} .
\end{eqnarray}
In the following we truncate the expansion of the quark propagator to the
ground state eigen mode:
\begin{eqnarray}\label{quark_propagator_ground}
iG_\psi(x,y) \to iG_0(x,y) \doteq u_0(\vec{x}) \, \bar u_0(\vec{y}) \,
e^{-i{\cal E}_\alpha(x_0-y_0)} \, \theta(x_0-y_0),
\end{eqnarray}
For $K$-mesons we use the free Feynman propagator for a boson field with
\begin{eqnarray}
i\Delta_K(x-y)=\int\frac{d^4k}{(2\pi)^4i}
\frac{\exp[-ik(x-y)]}{M_K^2 - k^2 - i\epsilon}.
\end{eqnarray}
Below, for transparency, we present the analytical expressions
for the strange nucleon form factors obtained in the PCQM.

a) The meson-cloud diagram (MC) (Fig.1a) results in:
\eq
\hspace*{-.5cm}
G_E^{\rm s}(Q^2)\bigg|_{MC}  &=& - \frac{27}{200} \,
\biggl(\frac{g_A}{\pi F}\biggr)^2
\int\limits_0^\infty dp \, p^2 \, \int\limits_{-1}^1 dx \,
(p^2+p\sqrt{Q^2}x) \, {\cal F}_{\pi NN}(p^2,Q^2,x) \,
t_E^{MC}(p^2,Q^2,x), \\
\hspace*{-.5cm}G_M^{\rm s}(Q^2)\bigg|_{MC}  &=& - \frac{9}{200} \, m_N \,
\biggl(\frac{g_A}{\pi F}\biggr)^2 \int\limits_0^\infty dp \, p^4 \,
\int\limits_{-1}^1 dx (1-x^2) \,\, {\cal F}_{\pi NN}(p^2,Q^2,x) \,
t_M^{MC}(p^2,Q^2,x),
\en
where
\eq
{\cal F}_{\pi NN}(p^2,Q^2,x)&=&F_{\pi NN}(p^2) \,
F_{\pi NN}(p^2+\Delta) , \hspace*{.5cm} \Delta = Q^2+2p\sqrt{Q^2}x ,
\nonumber\\\nonumber\\
t_E^{MC}(p^2,Q^2,x)&=&\frac{1}{w_K(p^2) w_K(p^2 + \Delta)
[w_K(p^2) + w_K(p^2 + \Delta)]} , \nonumber\\
t_M^{MC}(p^2,Q^2,x)&=&\frac{1}{w_K^2(p^2) w_K^2(p^2 + \Delta)}.
\en
with $\omega_K(t) = \sqrt{M_K^2 + t}$. 
Here $F_{\pi NN}(p^2)$ is the $\pi NN$ form factor normalized to unity
at zero recoil. For the {\it Gaussian} form of the single-quark wave 
function it is given by \cite{PCQM1}
\begin{eqnarray}\label{F_piNN}
F_{\pi NN}(p^2) = \exp\biggl(-\frac{p^2R^2}{4}\biggr) \biggl\{ 1 \, + \,
\frac{p^2R^2}{8} \biggl(1 \, - \, \frac{5}{3g_A}\biggr)\biggr\} .
\end{eqnarray}

b) The vertex-correction diagram (VC) (Fig.1b) contributes:
\eq
G_{E(M)}^{\rm s}(Q^2)\bigg|_{VC} = G_{E(M)}^p(Q^2)\bigg|_{3q}^{LO} \,
\cdot \, \frac{3}{200} \, \biggl(\frac{g_A}{\pi F}\biggr)^2 \,
\int\limits_0^\infty dp \, p^4 \, \frac{F_{\pi NN}^2(p^2)}{w^3_K(p^2)}
\, \cdot \, t_{E(M)}^{VC},
\en
\eq
G_A^{\rm s}(Q^2) \equiv G_A^{\rm s}(Q^2)\bigg|_{VC} = G_A(Q^2) \,
\cdot \, \frac{9}{1000} \, \biggl(\frac{g_A}{\pi F}\biggr)^2 \,
\int\limits_0^\infty dp \, p^4 \, \frac{F_{\pi NN}^2(p^2)}{w^3_K(p^2)},
\en
where $t_E^{VC} = 9$ and $t_M^{VC} = -1$; $G_A(Q^2)=g_A F_{\pi NN}(Q^2)$
is the  axial nucleon form factor; $G_{E(M)}^p(Q^2)\bigg|_{3q}^{LO}$
are the proton charge and magnetic form factors calculated at
leading-order~(LO)~\cite{PCQM1}
\begin{eqnarray}\label{GEP_3q}
G_E^p(Q^2)\bigg|_{3q}^{LO} &=& \exp\biggl(-\frac{Q^2R^2}{4}\biggr)
\biggl(1 - \frac{\rho^2}{1+\frac{3}{2}\rho^2} \frac{Q^2R^2}{4} \biggr),\\
G_M^p(Q^2)\bigg|_{3q}^{LO} &=& \exp\biggl(-\frac{Q^2R^2}{4}\biggr)
\frac{2m_N \rho R}{1+\frac{3}{2}\rho^2} .
\end{eqnarray}
For the case of the strange axial nucleon form factor only the 
VC diagram (Fig.1b) contributes at the order of accuracy we are working 
in. Combining the contributions of both diagrams, the expressions for 
the static characteristics (the strange magnetic moments, radii and 
axial charge of the nucleon) are given by
\eq
\mu^{\rm s}&=& - \, \frac{3}{200} \,
\biggl(\frac{g_A}{\pi F}\biggr)^2 \, \int\limits_0^\infty
\frac{dp \, p^4}{w_K^3(p^2)} \,\, F_{\pi NN}^2(p^2)\biggl[ \mu_p^{LO} \,
+ \, \frac{4m_N}{w_K(p^2)} \biggr] , \\
g^{\rm s}_A&=& - \, \frac{27}{4000} \,
\frac{\displaystyle{g_A}}{\displaystyle{1+\frac{5}{3g_A}}} \,
\biggl(\frac{g_A}{\pi F}\biggr)^2 \, \int\limits_0^\infty
\frac{dp \, p^4}{w_K^3(p^2)} \,\, F_{\pi NN}^2(p^2) , \\
<r^2>^{\rm s}_E &=& \frac{27}{200} \,
\biggl(\frac{g_A}{\pi F}\biggr)^2 \, \int\limits_0^\infty \,
\frac{dp \, p^4}{w_K^3(p^2)} \biggl[ F_{\pi NN}^2(p^2) \biggl(
<r^2_E>^{P}_{LO} \, - \, \frac{5}{2}
\frac{p^2 + 3m_K^2(p^2)}{w_K^4(p^2)} \biggr) \nonumber\\
&-& (F_{\pi NN}^\prime(p^2))^2 \biggr] , \\
<r^2>^{\rm s}_M &=& - \frac{9}{125} \,
\biggl(\frac{g_A}{\pi F}\biggr)^2 \, \int\limits_0^\infty \,
\frac{dp \, p^4}{w_K^3(p^2)} \biggl[ F_{\pi NN}^2(p^2)
\biggl( \frac{5}{16}R^2 \cdot \mu_p^{LO} \, + \,
\frac{p^2 + 3m_K^2(p^2)}{w_K^4(p^2)} \biggr) \nonumber\\
&+& \frac{1}{2} \, (F_{\pi NN}^\prime(p^2))^2 \biggr],
\en
where $F_{\pi NN}^\prime(p^2) = dF_{\pi NN}(p^2)/dp$. In the analytical 
expressions the strange nucleon properties are related to the leading 
order nonstrange results of the charge radius of the proton 
$<r^2_E>^{P}_{LO}$ of Eq. (\ref{rad_LO}) and the proton magnetic 
moment $\mu_p^{LO}$ given by 
\begin{eqnarray}
\mu_p^{LO} = G_M^p(0)\bigg|_{3q}^{LO} =
\frac{2m_N \rho R}{1+\frac{3}{2}\rho^2} \, . 
\end{eqnarray}
Our results for the static characteristics of the nucleon associated 
with the vector current are listed in Table I. A comparison with other 
theoretical predictions is also included. For completeness, we also give 
our prediction for the leading strange charge coefficient 
$\rho^{\rm s}$, which is defined as 
\eq
\rho^{\rm s} = \frac{dG_E^{\rm s}(\tau)}{d\tau}
\bigg|_{\tau=\frac{Q^2}{4m_N^2} = 0} \,\,\, .
\en
Note that $\rho^{\rm s}$ is also expected to be measured by the HAPPEX II
experiment \cite{HAPPEXII}. The two strange charge distribution 
parameters $<r^2>^{\rm s}_E$ and $\rho^{\rm s}$ are related in the PCQM 
as
\eq
\rho^{\rm s} \, = \, - \frac{2}{3} \, m_N^2 \, <r^2>^{\rm s}_E.
\en
As we mentioned before, the error bars in our theoretical results are 
due to a variation of the range parameter $R$ of the quark wave function 
from 0.55 fm to 0.65 fm. Taking also into account the other models, 
predictions cover a large range of values. In earlier 
calculations \cite{Leinweber_QCD}-\cite{Mathur},\cite{Jaffe}-\cite{Cohen}, 
\cite{Park} the strange magnetic moment is favored to be negative,  
recent calculations \cite{Silva,Hong}, however, give positive values 
for $\mu^{\rm s}$. Our value of $\mu^{\rm s}$ is relatively small and 
negative, close to the one of Ref. \cite{Glozman,Hannelius}. Comparing our 
predicted values to HBChPT, though the central values of the quantities are 
quite different, the results are consistent if we take the error bars into 
account. For example, the strange magnetic moment and charge radius in 
the perturbative chiral quark model are $-0.048 \pm 0.012$ and 
$-0.011 \pm 0.003$~fm$^2$, and the corresponding values in HBChPT 
are $0.18 \pm 0.34$ and $0.05 \pm 0.09$~fm$^2$~\cite{Hemmert2}. 
For comparison, we also present the results generated by Lattice QCD 
calculations: we indicate a calculation in the quenched 
approximation~\cite{Dong}, which was recently improved in~\cite{Mathur}, 
and one, where the extrapolation scheme takes careful consideration 
of chiral symmetry constraints~\cite{Leinweber}. 

As for the leading strange charge coefficient $\rho^{\rm s}$, the spread of 
predictions of various models is quite large, ranging from $ - 2.93$ to 
$3.06$. Our value for $\rho^{\rm s}$ is positive and small: 
$0.17\pm 0.04$. Our results for the strange radii are also small when 
compared to other approaches. Because of the simple relationship between 
$\rho^{\rm s}$ and the strange charge radius $<~r^2>_E^{\rm s}$, a measurement 
of $\rho^{\rm s}$ also determines $<r^2>_E^{\rm s}$. A new $^4$He experiment 
was proposed to measure $\rho^{\rm s}$ at $Q^2\rightarrow 0$ to an accuracy 
of $\delta\rho^{\rm s} = \pm 0.5$ which could be capable of yielding a reasonable 
experimental value for $\rho^{\rm s}$ \cite{HAPPEXII}. In Ref. \cite{HAPPEXII} 
the authors obtain a preliminary value of \eq\label{rho_mu} 
\rho^{\rm s}+2.9\mu^{\rm s} = 0.67 \pm 0.41 \pm 0.30
\en
from the current HAPPEX data \cite{Happex2}. In Table I we indicate our 
prediction for $\rho^{\rm s}+\mu_p\mu^{\rm s}$ of 0.05 at central 
values of $\rho^{\rm s}$, $\mu_p$ and $\mu^{\rm s}$, which is consistent 
with Eq. (\ref{rho_mu}). The values obtained from lattice calculations and the 
Skyrme model, where the strange magnetic moment is negative as in the PCQM, 
also compare reasonably to the value of Eq. (\ref{rho_mu}). The results of 
dispersive approaches \cite{Jaffe,Forkel} and the NJL soliton 
model \cite{Weigel} disagree with the current experimental limits. 

We also calculate the combinations of strange form factors measured by 
SAMPLE and HAPPEX at finite $Q^2$ with our predictions as:
$G_{\rm SAMPLE}^{\rm s}(Q^2_{\rm S})=-(3.7\pm 1.2) \times 10^{-2}$ and
$G_{\rm HAPPEX}^{\rm s}(Q^2_{\rm H})=(1.8\pm 0.3) \times 10^{-3}$.
The prediction for the MAMI A4 experiment is
$G_{\rm MAMI}^{\rm s}(Q^2_{\rm M})=(2.9 \pm 0.5) \times 10^{-4}$.
In Table II we summarize the theoretical results for the respective 
experiments obtained within different theoretical approaches.
The values of our predictions for the HAPPEX and MAMI experiments are 
quite small. This is, because in the PCQM we have a cancellation between 
the strange charge form factor, which is positive and the magnetic one, 
which turns out to be negative, as will be shown below. In HBChPT, the 
central values of the SAMPLE and HAPPEX experiments are fitted by an  
appropriate choice of the low-energy couplings (LECs) \cite{Hemmert2}. 
We also determine the strange axial charge of the nucleon with 
\begin{eqnarray}
g_A^{\rm s} = - 0.0052 \pm 0.0015 \,\,\, . 
\end{eqnarray}
Out prediction has the correct sign when compared to the present direct 
estimate of this quantity from deep inelastic scattering 
experiments \cite{Alberico}: $g_A^{\rm s} = - 0.12 \pm 0.03$, 
but differs quantitatively. 

Next we discuss the momentum dependence of the strange form factors for 
the central value of our size parameter $R=0.6$ fm. In Fig.2, we plot 
the $Q^2$ dependence of the nucleon charge strange form factor, which is 
positive for all $Q^2$ and has maximum at $Q^2 \sim 0.5$ GeV$^2$. The 
behavior of $G_E^s(Q^2)$ is close to the one calculated in \cite{Dubnicka} 
and similar to the results obtained in Lattice QCD \cite{Dong} and the chiral 
soliton quark model \cite{Silva}. All these calculations give a positive 
strange form factor $G_E^s$. Our result is also consistent with that 
of Ref. \cite{Hemmert2} considering the uncertainty of the parameters. The 
chiral quark model of Ref. \cite{Hannelius}, however, obtains a form factor 
$G_E^{\rm s}$ which is negative and decreases with the increasing $Q^2$. 

In Fig.3, we show the momentum dependence of the strange magnetic form 
factor $G_M^{\rm s}(Q^2)$ which is similar to the prediction of Lattice 
QCD \cite{Dong}. Both results give a negative strange form factor 
$G_M^{\rm s}(Q^2)$. Though our results for $G_E^{\rm s}$ is close to 
Refs. \cite{Dubnicka} and \cite{Silva}, the predictions for $G_M^{\rm s}$ 
are different, in that in latter references a positive value for 
$G_M^{\rm s}$ is obtained for the indicated $Q^2$ range. 
In Ref. \cite{Hemmert2}, both positive and negative value of $G_M^{\rm s}$ 
are possible due to the uncertainty of the parameters. 

In Fig.4 and 5 we plot the momentum dependence of the Dirac 
$F_1^{\rm s}(Q^2)$ and Pauli $F_2^{\rm s}(Q^2)$ strange form factors of 
the nucleon. The behavior of $F_1^{\rm s}$ is close to that of 
$G_E^{\rm s}$, while $F_2^{\rm s}$ and $G_M^{\rm s}$ have similar momentum 
dependence. Finally, in Fig.6 we plot the $Q^2$-dependence of the normalized 
strange axial nucleon form factor $G_A^{\rm s}(Q^2)/G_A^{\rm s}(0)$.

\section{Summary} 

In summary, we investigate the strange vector (charge and magnetic) and 
axial form factors of the nucleon in the perturbative chiral quark model. 
We determined the momentum dependence of the strange form factors, radii and 
magnetic moments at one loop level. Our results for the strange form factors 
and the leading strange charge coefficient are consistent with the SAMPLE and 
HAPPEX experiments. The strange electric form factor $G_E^{\rm s}$ is positive 
with a maximum at about $Q^2 \sim 0.5$ GeV$^2$. The strange magnetic form 
factor $G_M^{\rm s}$ is negative and its absolute value increases with $Q^2$. 
The signs of the discussed quantities are stable with respect to changes in 
the model parameter. Absolute values for both $G_E^{\rm s}$ and $G_M^{\rm s}$ 
are small in this model, consistent with currently available data. An accurate 
determination of the strangeness form factors at moderate $Q^2$ will provide 
a sensitive test for the modeling of sea quark contributions in the 
low-energy description of the nucleon and hence for the implementation 
of chiral symmetry. 

\vspace*{.5cm}

{\bf Acknowledgements}

\vspace*{.1cm}

\noindent
P. Wang would like to thank the Institute for Theoretical Physics,
University of T\"ubingen for their hospitality. This work was
supported by the Alexander von Humboldt Foundation and by the Deutsche
Forschungsgemeinschaft (DFG) under contracts FA67/25-1, GRK683.

\newpage
\begin{table}[t]
\caption{Static strange characteristics associated with the vector current.}
\begin{center}
\begin{tabular}{|c|c|c|c|c|c|}
Approach& $\mu^{\rm s}$ (n.m.) & $<r^2>_E^{\rm s} ({\rm fm}^2)$ &
$<r^2>_M^{\rm s} ({\rm fm}^2)$ & $\rho^{\rm s}$
         & $\rho^{\rm s} + \mu_p \mu^{\rm s}$ \\
\hline\hline
QCD equalities \cite{Leinweber_QCD} & $ - 0.75 \pm 0.30$ & & & & \\
Lattice QCD \cite{Leinweber} & $ - 0.16 \pm 0.18$ & & & &\\
Lattice QCD \cite{Dong} & $ - 0.36 \pm 0.20$ & $ - 0.16 \pm 0.06$ &
& $2.02 \pm 0.75$ & $1 \pm 0.75$ \\
Lattice QCD \cite{Mathur} & $-0.28 \pm 0.10$ & & & &\\
\hline
HBChPT \cite{Hemmert2} & $0.18 \pm 0.34$
& $0.05\pm 0.09$ & $- 0.14$ & &\\ \hline
\hline
Poles \cite{Jaffe}   & $ -0.31\pm 0.09$ & $0.14 \pm 0.07$
& & $- 2.1$ & $- 2.97$\\
Poles \cite{Forkel} & $ -0.185 \pm  0.075$ &  $0.14 \pm 0.06$
& & $- 2.93$ & $ - 3.60$\\
Poles \cite{Hammer} & $ -0.24 \pm  0.03$ & & & &\\
\hline
Kaon loop \cite{Musolf} & $ -0.355 \pm  0.045$ & $ - 0.0297 \pm 0.0026$
& & &\\
Kaon loop + VMD \cite{Cohen}  & $ - 0.28 \pm 0.04$
&  $ -0.0425 \pm  0.0026$ & & &\\
\hline
Skyrme model \cite{Park} & $ - 0.13$ &  $ - 0.11$ & & 1.64 & 1.27\\
NJL soliton model \cite{Weigel} & $ 0.10 \pm 0.15$ & $ - 0.15 \pm 0.05$ 
& & 3.06 & 2.92\\
\hline
$\chi$QSM \cite{Silva} & 0.115 & $- 0.095$ & 0.073 & & \\
\hline
CBM \cite{Hong} & 0.37 & & & & \\
CQM \cite{Glozman}   & $\sim -0.05$ & & & & \\
CQM \cite{Hannelius} & $- 0.046$ & & $\sim 0.02$ & & \\ \hline
PCQM & $ - 0.048 \pm 0.012$ & $- 0.011 \pm 0.003$ & $ - 0.024 \pm 0.003$
& $0.17 \pm 0.04$ & 0.05\\
\end{tabular}
\end{center}
\end{table}

\begin{table}
\caption{Predictions for SAMPLE, HAPPEX and MAMI experiments in 
comparison to current experimental results.}
\begin{center}
\begin{tabular}{|c|c|c|c|}
Approach&  $G_{\rm SAMPLE}^s(Q^2_{\rm S})$
        &  $G_{\rm HAPPEX}^s(Q^2_{\rm H})$
        & $G_{\rm MAMI}^s(Q^2_{\rm M})$ \\
\hline\hline
HBChPT \cite{Hemmert2}
& 0.23 $\pm$ 0.44 & $ 0.023 \pm 0.048$ & $ 0.007 \pm 0.127$  \\
\hline
$\chi$QSM \cite{Silva} & 0.09& $0.087 \pm 0.016$ & $0.141 \pm 0.033$\\
\hline
CQM \cite{Hannelius} & -0.06 & -0.08 & \\ \hline
PCQM & $-(3.7\pm 1.2) \times 10^{-2}$
     & $ (1.8\pm 0.3) \times 10^{-3}$
     & $ (2.9 \pm 0.5)\times 10^{-4}$ \\ \hline 
Experiment \cite{Sample2,Happex2} & $0.14 \pm 0.29 \pm 0.31$ 
     & $0.025 \pm 0.020 \pm 0.014$ &\\
\end{tabular}
\end{center}
\end{table}

\begin{figure}

{\bf Fig. 1:} Diagrams contribution to the strange nucleon form factors:

\hspace*{1.4cm} (a) meson cloud diagram and (b) vertex correction
diagram.
\bigskip

{\bf Fig. 2:} Nucleon strange charge form factor $G_E^{\rm s}(Q^2)$.
\bigskip

{\bf Fig. 3:} Nucleon strange magnetic form factor $G_M^{\rm s}(Q^2)$.
\bigskip

{\bf Fig. 4:} Nucleon strange Dirac form factor $F_1^{\rm s}(Q^2)$.
\bigskip

{\bf Fig. 5:} Nucleon strange Pauli form factor $F_2^{\rm s}(Q^2)$.
\bigskip

{\bf Fig. 6:} Normalized nucleon strange axial form factor
$G_A^{\rm s}(Q^2)/G_A^{\rm s}(0)$.

\end{figure}

\begin{figure}
\centering{\
\epsfig{figure=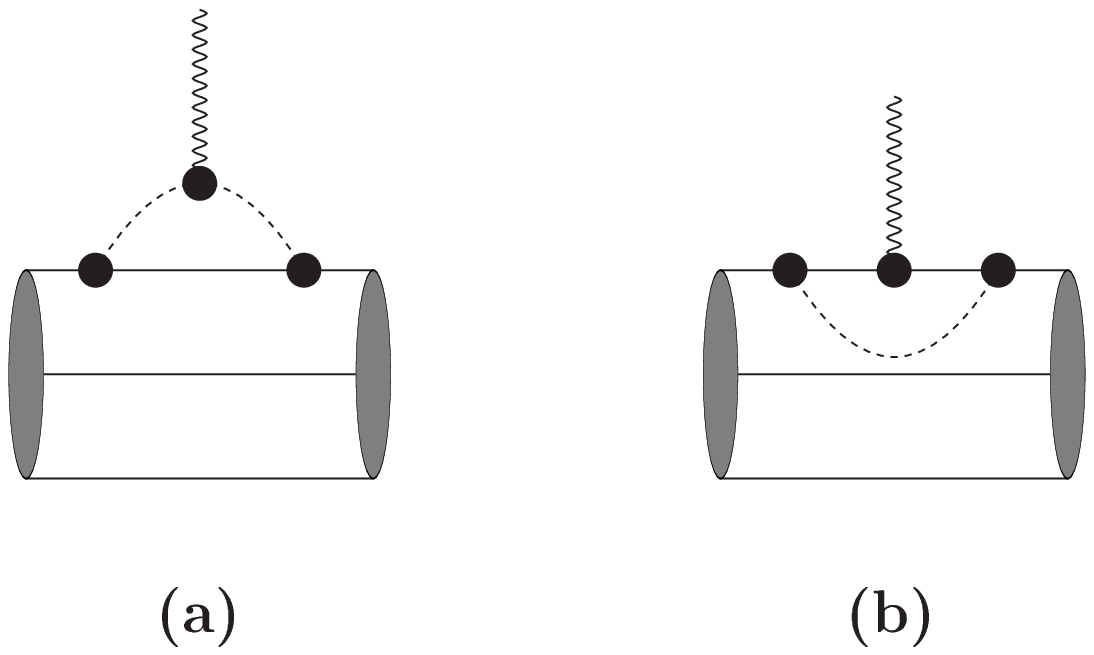,height=21cm}}
\end{figure}

\vspace*{-14cm}
\centerline{\bf Fig.1}

\begin{figure}
\centering{\
\epsfig{figure=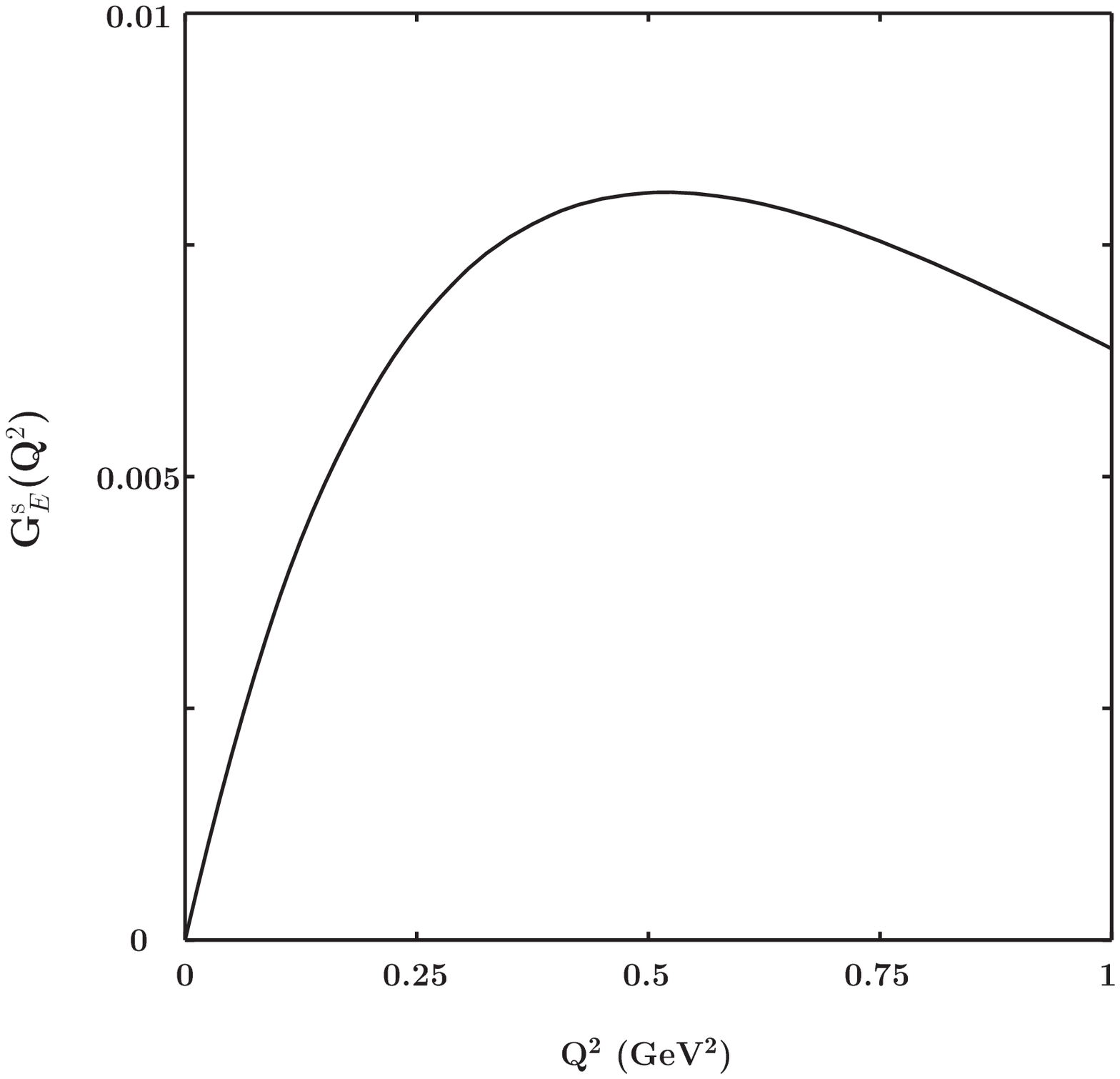,height=21cm}}
\end{figure}

\vspace*{-1cm}
\centerline{\bf Fig.2}

\begin{figure}
\centering{\
\epsfig{figure=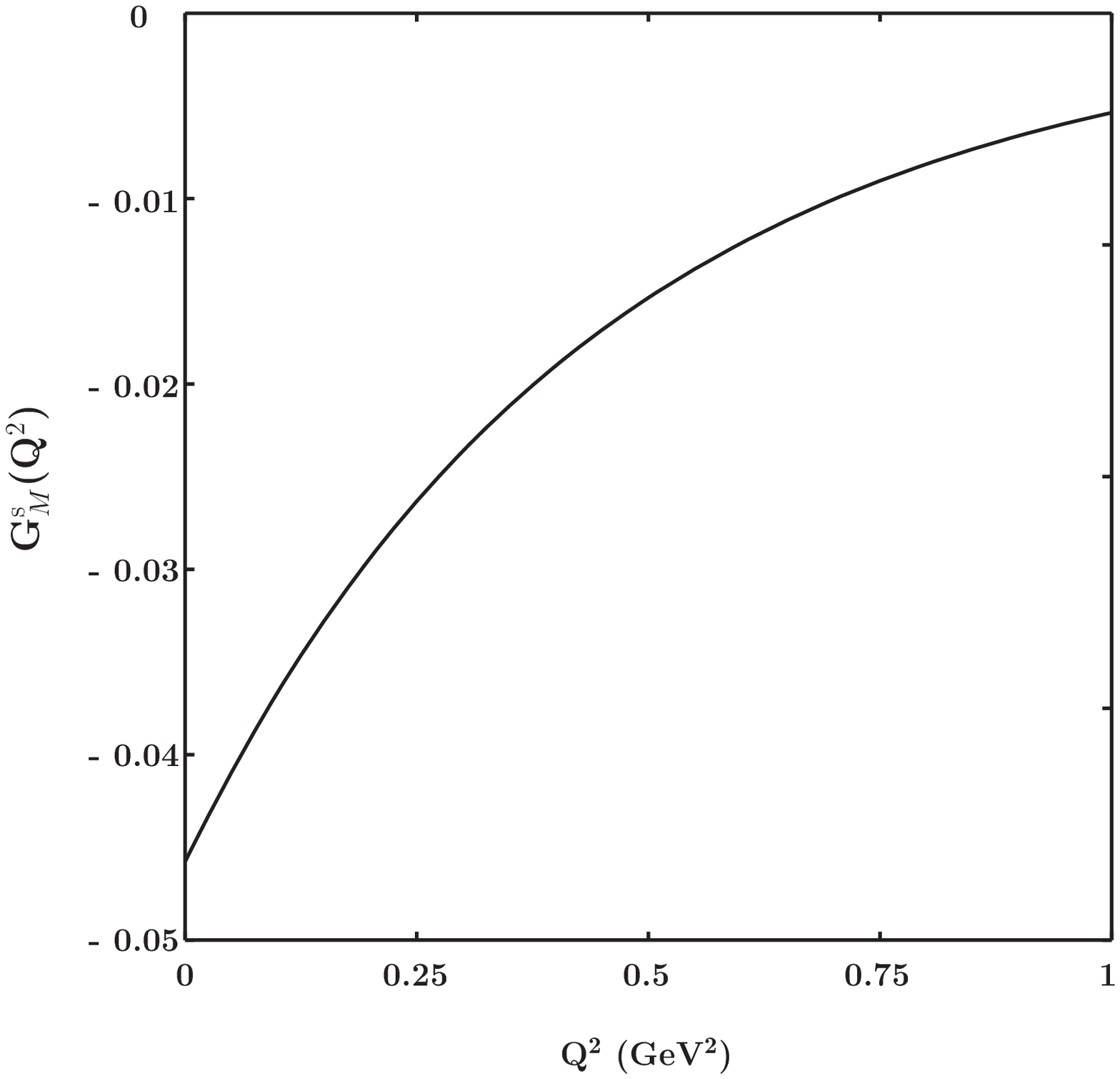,height=21cm}}
\end{figure}

\vspace*{-1cm}
\centerline{\bf Fig.3}

\begin{figure}
\centering{\
\epsfig{figure=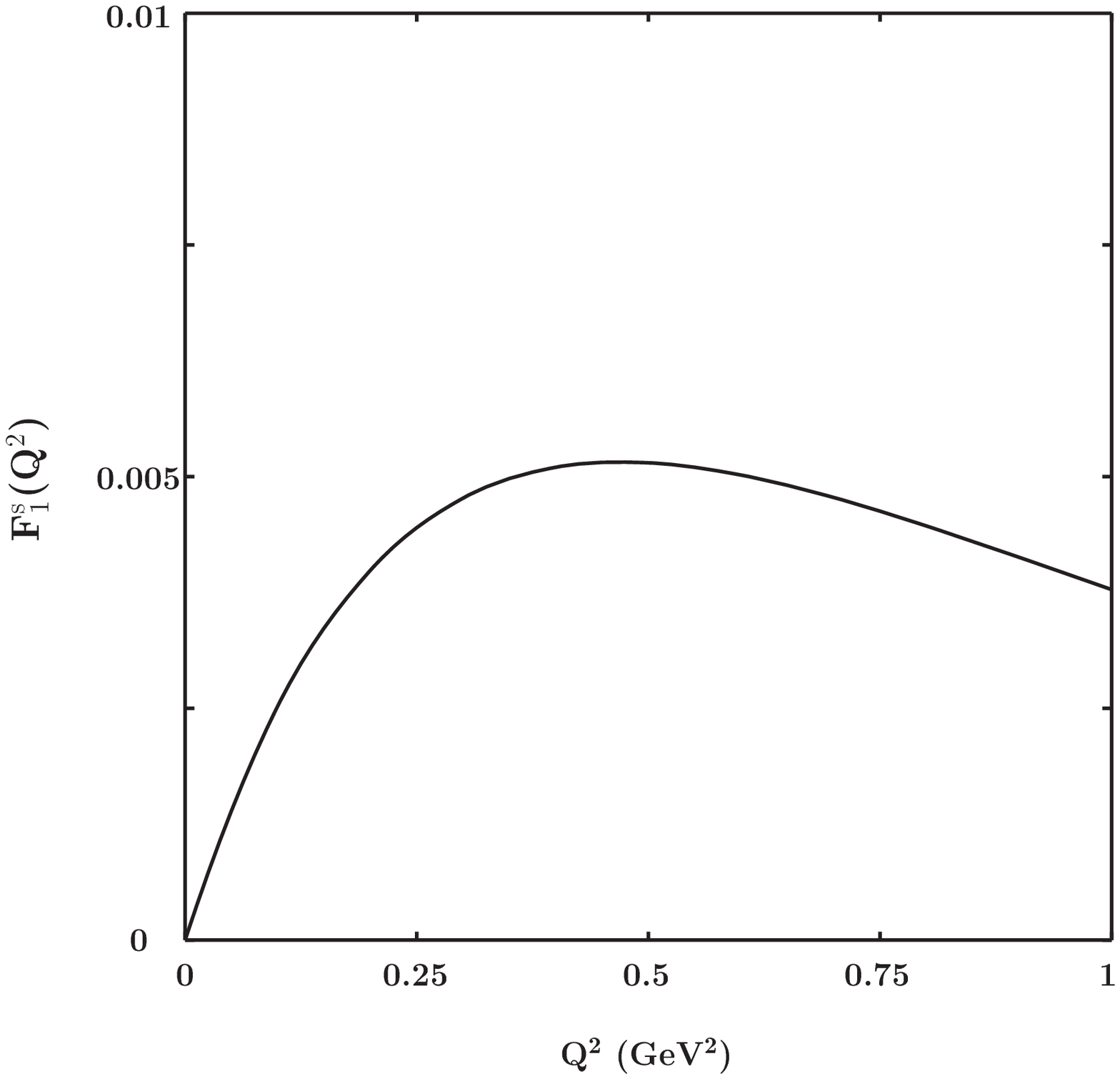,height=21cm}}
\end{figure}

\vspace*{-1cm}
\centerline{\bf Fig.4}

\begin{figure}
\centering{\
\epsfig{figure=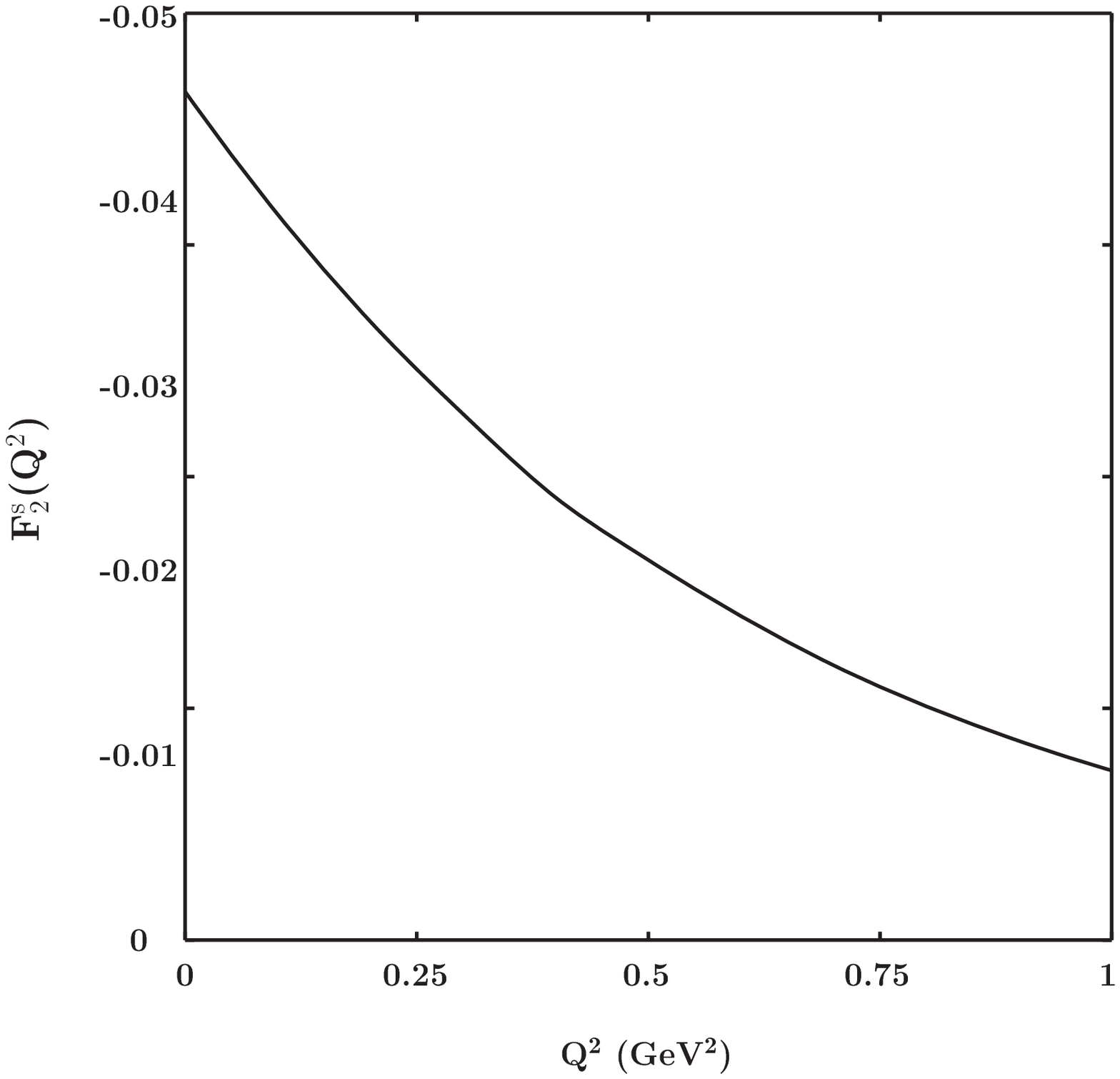,height=21cm}}
\end{figure}

\vspace*{-1cm}
\centerline{\bf Fig.5}

\begin{figure}
\centering{\
\epsfig{figure=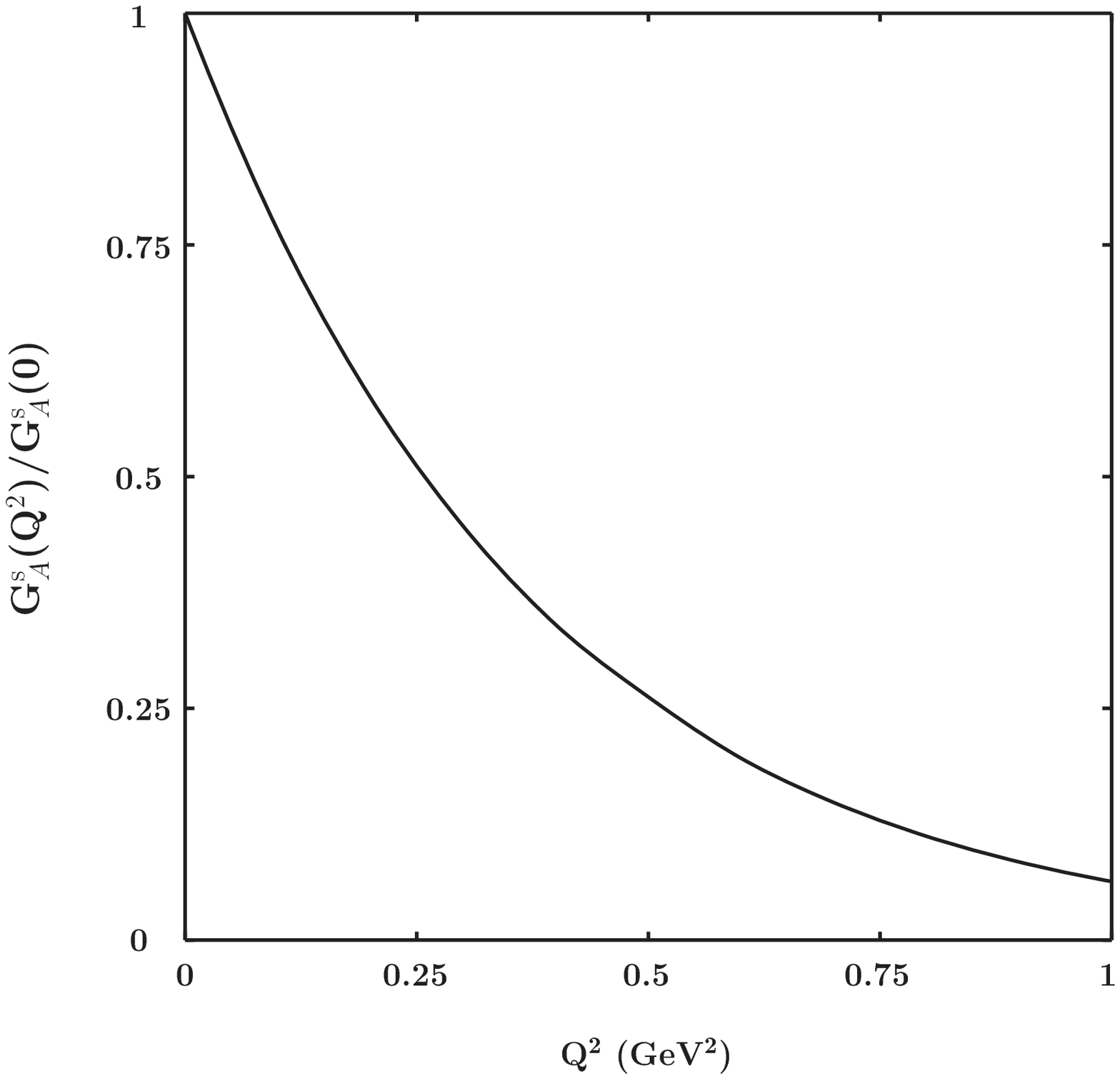,height=21cm}}
\end{figure}

\vspace*{-1cm}
\centerline{\bf Fig.6}

\end{document}